\documentclass[12pt]{iopart}
\usepackage{iopams}

\usepackage{color}
\usepackage{graphicx}

\usepackage[update,prepend]{epstopdf}

\usepackage{bm}

\usepackage{hyperref}

\newcommand{\bp}{\bm{p}}
\newcommand{\bq}{\bm{q}}
\newcommand{\bx}{\bm{x}}
\newcommand{\by}{\bm{y}}
\newcommand{\bX}{\bm{X}}
\newcommand{\bY}{\bm{Y}}

\newcommand{\avg}[1]{\left\langle{#1}\right\rangle}

\newcommand{\ox}{\overline{x}}
\newcommand{\oy}{\overline{y}}
\newcommand{\obx}{\overline{\bm{x}}}
\newcommand{\oby}{\overline{\bm{y}}}
\newcommand{\obp}{\overline{\bm{p}}}
\newcommand{\obq}{\overline{\bm{q}}}

\newcommand{\tx}{\tilde{x}}
\newcommand{\ty}{\tilde{y}}
\newcommand{\tbx}{\tilde{\bm{x}}}
\newcommand{\tby}{\tilde{\bm{y}}}

\newcommand{\hbzeta}{\hat{\bm{\zeta}}}
\newcommand{\hbvarphi}{\hat{\bm{\varphi}}}

\newcommand{\bone}{\bm{1}}

\newcommand{\suone}{^{(1)}}
\newcommand{\sutwo}{^{(2)}}
\newcommand{\sui}{^{(i)}}

\newcommand{\suonetrans}{^{(1)^T}}
\newcommand{\sutwotrans}{^{(2)^T}}

\begin{document}

\title{Effects of noise on convergent game learning dynamics}

\author{James BT Sanders$^1$, Tobias Galla$^1$ and Jonathan L Shapiro$^{2,3}$}

\address{
$^1$Theoretical Physics, School of Physics and Astronomy, The University of Manchester, Manchester M13 9PL, United Kingdom\\
$^2$School of Computer Science, The University of Manchester, Manchester M13 9PL, United Kingdom\\
$^3$Centre for Interdisciplinary Computational and Dynamical Analysis (CICADA), The University of Manchester, Manchester M13 9PL, United Kingdom
}

\begin{abstract}
  We study stochastic effects on the lagging anchor dynamics, a reinforcement learning algorithm used to learn successful strategies in iterated games, which is known to converge to Nash points in the absence of noise.  The dynamics is stochastic when players only have limited information about their opponents' strategic propensities.   The effects of this noise are studied analytically in the case where it is small but finite, and we show that the statistics and correlation properties of fluctuations can be computed to a high accuracy. We find that the system can exhibit quasicycles, driven by intrinsic noise. If players are asymmetric and use different parameters for their learning, a net payoff advantage can be achieved due to these stochastic oscillations around the deterministic equilibrium.
\end{abstract}

\ead{\mailto{James.Sanders-2@postgrad.manchester.ac.uk}, \mailto{Tobias.Galla@manchester.ac.uk}, \mailto{jls@cs.man.ac.uk}}


\section{Introduction}
Competitive situations in biology, the social sciences, and computer science are often modelled using the formalism of game theory~\cite{Neumann1944}.  In these models, two or more players each choose between a set of actions, and each player then receives a reward (or `payoff'), depending on their own action and those of the other players.  Traditionally, games are classified in terms of outcomes one would expect for perfectly rational players, so-called Nash equilibria~\cite{Nash1950, Nash1951}.  These are points in strategy space such that no player can improve their expected payoff by unilaterally changing their own behaviour. In many simple games, it is straightforward to compute such equilibrium points. Typical equilibrium strategies will be probabilistic---these are known as `mixed strategies', while a deterministic strategy is described as `pure'.  In games with a large number of actions to choose between, there may be many equilibria, or even none at all if there are infinitely many actions.  Even when a unique optimal solution to a game is available, the players may not have enough information or computational capacity to identify it. To model situations where an optimal strategy does not exist, or where the players are unable to adopt it, learning algorithms have been introduced. These describe scenarios in which players repeatedly play a game, modifying their strategies each time to try and maximize the payoffs they receive~\cite{Sato2002, Sato2003, Camerer2003, Ho2007, Macy2002, Young2004, Fudenberg1998}. The purposes of these algorithms vary.  Some are attempts to model natural or social systems.  Other learning algorithms are used in decision-making software, optimization, and more generally in machine learning, a branch of computer science concerned with developing algorithms to improve behaviour using empirical information~\cite{Alpaydin2004}. In our work we do not aim to model the psychological processes underpinning human learning, but instead focus on a class of machine-learning algorithms proposed by Dahl~\cite{Dahl2005}, the so-called `lagging anchor dynamics'. Before we define the exact details of this dynamics it is useful to briefly outline some of the basic general principles of game learning.

Learning algorithms typically require a player to play a mixed strategy, iteratively modifying the probabilities of playing each action depending on the outcomes of past games.  In general, the player gradually increases the probability of playing actions that would achieve higher payoffs against their opponent's current strategy, while decreasing the probability of those that would perform poorly~\cite{Fudenberg1998,  Gintis2000}.  One well-studied adaptation mechanism in machine learning is reinforcement learning. In this class of dynamics, the only information available to a player is the relative success or failure of previous actions~\cite{Alpaydin2004,Sutton1998}.

In game learning the expected payoff can be viewed as a function of the players' mixed strategies, and each player wishes to maximize their own payoff, so that reinforcement learning can be seen as an optimization problem.  One well-known optimization method is gradient ascent, an iterative scheme in which the maximum of a function is approached by repeatedly taking small steps in the direction in which the value of the function increases most quickly. If $\bm{x}^{(0)}$ is an initial estimate of the maximum of the function $f(\cdot)$, then gradient ascent proceeds according to the recurrence relation
\begin{equation} \label{gradientascenteqn}
	\bm{x}^{(n + 1)} = \bm{x}^{(n)} + \kappa \nabla f(\bm{x}^{(n)}),
\end{equation}
where the step size $\kappa$ is a small positive constant~\cite{Chong2008}.  In the context of games, each player views their expected payoff as a function of their mixed strategy, imagining the other players' strategies are constant, and carries out one step of the gradient ascent method.  This is sometimes known as `simultaneous gradient ascent'~\cite{Singh2000}.  If each player has complete knowledge of their opponents' mixed strategies, gradient ascent can be used deterministically.  If not, the players must estimate their opponents' strategies based on observations.  This introduces noise into the system, as the opponents' actions are drawn probabilistically from an underlying mixed strategy, which is unknown to the observer.

In a game with a single Nash equilibrium point, it is often argued that rational players would choose the equilibrium strategies, in which case, it is desirable that a learning algorithm should find these strategies.  The standard simultaneous gradient ascent algorithm is typically able to converge to pure equilibrium strategies, but not mixed strategies~\cite{Dahl2005}.  A number of variations of the algorithm have been proposed to address this issue.  For example, in some versions the players attempt to predict their opponents' future strategies, and calculate the gradient of the payoff function based on these expected future strategies rather than the current ones. Another variation is `satisficing', in which each player maintains an `aspiration', and only updates their strategy if their expected payoff is lower than their current aspiration~\cite{Dahl2005}.

The lagging anchor algorithm is another modification of gradient ascent, in which each player maintains a long-term memory of past strategies---a `lagging anchor'---which is coupled to their current strategy.  This algorithm was introduced by Dahl~\cite{Dahl2005}, who concentrated on its deterministic behaviour, when each player knows precisely their opponents' current mixed strategies.  He was able to prove that the dynamics is able to converge to mixed equilibrium strategies in a broad class of games.  Butterworth and Shapiro~\cite{Butterworth2009} looked at the continuous-time limit of the system, deriving further results concerning the deterministic behaviour.  They also investigated two different situations in which the players have limited information, noting the appearance of stochastic quasicycles.

The main purpose of our work is to extend the analysis of stochastic lagging anchor learning.  We concentrate on situations where the intrinsic noise is small but finite, and examine analytically its effect upon the system. In particular, we investigate how stochasticity affects the stability of the dynamics, and whether players can exploit noise in the system to increase their own payoffs.

\section{Model definitions}
\subsection{General definitions}
Consider a finite, two-player game with $m$ actions available to player~$1$, and $n$ for player~$2$.  A particular game is defined by two $m \times n$ payoff matrices, $E\suone$ and $E\sutwo$, so that if player~$1$ chooses action $\alpha$, and player~$2$ chooses action $\beta$, the payoff to player $i$ is $E\sui_{\alpha \beta}$.

At any given time step $t$, the players have mixed strategies represented by two vectors $\bp(t)$ and $\bq(t)$, and one instance of the game is played.  Player~1 chooses action $\alpha$ with probability $p_\alpha(t)$, while player~$2$ chooses action $\beta$ with probability $q_\beta(t)$.  The corresponding payoffs are distributed to the players, who then update their strategies according to their learning method (this will be described below).

For simplicity, we initially focus on the so-called `matching pennies' game~\cite{Neumann1944}.  In this zero-sum game, each player has a coin and selects `heads' or `tails'.  If the players' choices match, player~$1$ keeps both pennies, otherwise player~$2$ keeps them both.  The payoff matrices for this two-action game are
\begin{equation*}
	E\suone = \pmatrix{1 & -1 \cr -1 & 1} = - E\sutwo.
\end{equation*}
The unique Nash equilibrium for this game is the pair of mixed strategies for which each player chooses heads or tails with equal probabilities.  For convenience, we write the mixed strategies as $\bp(t) = \left(\frac{1}{2} + x(t), \frac{1}{2} - x(t)\right)^T$ and $\bq(t) = \left(\frac{1}{2} + y(t), \frac{1}{2} - y(t)\right)^T$, so that the Nash equilibrium is at $x=y=0$, the expected payoff to player~$1$ is $\bp^T E\suone \bq = 4xy$, and the expected payoff to player~$2$ is $-4xy$.
\subsection{Gradient ascent}
Rephrasing the matching pennies game as an optimization problem, player~$1$ wishes to maximize $4xy$, and is able to vary $x$ in the interval $[-\frac{1}{2}, \frac{1}{2}]$.  Player~$2$ wants to maximize $-4xy$, and can vary $y$ over the same interval.  If each player applies the gradient ascent method~\eref{gradientascenteqn}, the update rules read
\begin{equation}\label{gradascentpennies}
\eqalign{
	x(t+1) &= x(t) + 4 \kappa_1 \ty(t), \cr
	y(t+1) &= y(t) - 4 \kappa_2 \tx(t),
}
\end{equation}
where $\kappa_1$ and $\kappa_2$ are the two players' gradient ascent step sizes, and can be thought of as learning rates.  In the event that $x$ or $y$ are taken out of their allowed intervals by these update rules, they are simply mapped back to the nearest allowed points.  Here we use the notation $\tx(t)$ to represent player~$1$'s estimate of player~$2$'s strategy at time $t$, and respectively for $\ty(t)$.  We assume the players calculate these estimates using a geometric discounted average of their opponents' previous choices as in~\cite{Butterworth2009}, giving update rules for $\tx$ and $\ty$,
\begin{equation*}
\eqalign{
	\tx(t+1)	= \tx(t) + \phi_1 \left( X(t) - \tx(t) \right),	\cr
	\ty(t+1)	= \ty(t) + \phi_2 \left( Y(t) - \ty(t) \right),
}
\end{equation*}
where $\phi_1$ and $\phi_2$ are constant parameters, and $X(t)$ and $Y(t)$ represent the pure strategies chosen by players~$1$ and~$2$, respectively, at time $t$, taking the value $+\frac{1}{2}$ for action~$1$, and $-\frac{1}{2}$ for action~$2$.  Deterministic mean field equations are recovered by setting $X = x$ and $Y = y$.\footnote{Another method suggested in~\cite{Butterworth2009} covers situations where the players cannot observe their opponents' actions, and must calculate gradient estimates using only observations of their own payoffs.  Using this method instead of opponent modelling appears to have little effect on the dynamics, except to increase the size of stochastic effects.}\footnote{Here, we follow Butterworth and Shapiro in updating the strategies at time $t$ based on the estimates from time $t-1$, which only include contributions from observations from time $t-2$ and earlier.  It could be argued that the players should include observations from time step $t-1$ in their strategy update at time $t$, but this would be unlikely to change the dynamics significantly.}

\subsection{Lagging anchor dynamics}
Although the Nash equilibrium ($x=y=0$) is a fixed point of the deterministic gradient ascent equations~\eref{gradascentpennies}, it is not stable and the system will typically not converge to this equilibrium. This is a general feature of mixed Nash equilibria under gradient-type strategy updates, as has been known for a long time (see, for example~\cite{Selten1991, Crawford1974}). Several mechanisms have been proposed to make equilibria stable under gradient-type updates, among them the lagging anchor algorithm we study in this work. The lagging anchor algorithm involves introducing an additional state variable for each player.  These variables, called `lagging anchors', are weighted averages of previously used mixed strategies. For matching pennies, the lagging anchor dynamics, with the same geometric discounted estimates, is
\begin{equation}\label{fullsystem}
\eqalign{
	x(t+1)		= x(t) + 4 \kappa_1 \ty(t) + \mu_1 \left( \ox(t) - x(t) \right), 	\cr
	y(t+1)		= y(t) - 4 \kappa_2 \tx(t) + \mu_2 \left( \oy(t) - y(t) \right), 	\cr
	\ox(t+1)	= \ox(t) + \nu_1 \left( x(t) - \ox(t) \right),										\cr
	\oy(t+1)	= \oy(t) + \nu_2 \left( y(t) - \oy(t) \right),										\cr
	\tx(t+1)	= \tx(t) + \phi_1 \left( X(t) - \tx(t) \right),										\cr
	\ty(t+1)	= \ty(t) + \phi_2 \left( Y(t) - \ty(t) \right),
}
\end{equation}
where the lagging anchors are given by $\obp(t) = (\frac{1}{2} + \ox(t), \frac{1}{2} - \ox(t))^T$ and $\obq(t) = (\frac{1}{2} + \oy(t), \frac{1}{2} - \oy(t))^T$. The introduction of the anchors does not change the location of the equilibrium of the deterministic dynamics, but it does stabilize it if the parameters are chosen appropriately. The variables $\mu_i$ and $\nu_i$ ($i=1,2$) are non-negative model parameters, whose interpretation will be detailed further below. The system of equations~\eref{fullsystem} is a generalization of that presented in~\cite{Butterworth2009}, which is recovered if $\kappa_1 = \kappa_2$, $\mu_1 = \mu_2 = \nu_1 = \nu_2$, and $\phi_1 = \phi_2$. As before we will implicitly assume that any variable leaving the interval $[-1/2,1/2]$ will be clipped and mapped onto the points $1/2$ or $-1/2$ respectively. This introduces an occasional nonlinearity into a dynamics whose deterministic behaviour would otherwise be linear.

The update rules~\eref{fullsystem} contain eight constant parameters.  The gradient ascent step sizes $\kappa_1$ and $\kappa_2$ can be thought of as learning rates, and must be small and positive for the dynamics to be capable of learning the equilibrium strategy.  The anchor parameters $\mu_i$ and $\nu_i$ determine the strength of coupling between the strategies and anchors.  A large $\nu_i$ corresponds to anchors that are pulled strongly towards the mixed strategies, and a large $\mu_i$ corresponds to strategies that are pulled strongly towards the anchors.  In particular, setting $\mu_i$ to zero or $\nu_i$ to one is equivalent to removing the lagging anchors altogether, and setting $\nu_i$ to zero will hold the anchors in their initial positions, permanently biasing the strategies towards these points. Finally, the modelling rates $\phi_1$ and $\phi_2$ represent the rates at which the players update their models of their opponents' strategies.  A value of $\phi_i=1$ is equivalent to the players discarding all previous information about their opponents' actions and assuming that their strategy is simply to play their most recently used action every game.  A value of $\phi_i$ between 0 and 1 will cause the opponent models to be averages of past behaviour, weighted according to a geometric discounting factor.

\subsection{Compact notation}
For convenience, we rewrite the full stochastic lagging anchor system~\eref{fullsystem} in matrix form, separating the deterministic and stochastic effects.  Since the observed pure strategies $X$ and $Y$ are unbiased estimates of the mixed strategies $x$ and $y$, respectively, we can write them in terms of mean-zero noise terms $\xi$ and $\chi$,
\begin{equation}\label{noisedef}
\eqalign{
	X(t) = x(t) + \xi(t),		\cr
	Y(t) = y(t) + \chi(t).
}
\end{equation}
More precisely, $\xi(t)$ takes the value $1/2-x(t)$ with probability $1/2+x(t)$, and the value $-1/2-x(t)$ with probability $1/2-x(t)$, and similar for $\chi(t)$. The state of the system at any given time is fully described by the vector $\bzeta = \pmatrix{x	&	y	&	\ox & \oy & \tx & \ty}^T$. The update rules~\eref{fullsystem} can be written in matrix form as
\begin{equation}\label{fullsystemmatrix}
	\bzeta(t+1) = J \bzeta(t) + \bvarphi(t),
\end{equation}
where the column vector $\bvarphi = \pmatrix{0 & 0 & 0 & 0 & \phi_1 \xi & \phi_2 \chi}^T$ contains the noise terms, and the constant matrix $J$ is
\begin{equation*}
	J =
	\pmatrix{
		1 - \mu_1		&	0						&	\mu_1			&	0					&	0							&	4 \kappa_1	\cr
		0						&	1 - \mu_2		&	0					&	\mu_2			&	- 4 \kappa_2	&	0						\cr
		\nu_1				&	0						&	1 - \nu_1	&	0					&	0							&	0						\cr
		0						&	\nu_2				&	0					&	1 - \nu_2	&	0							&	0						\cr
		\phi_1			&	0						&	0					&	0					&	1 - \phi_1		&	0						\cr
		0						&	\phi_2			&	0					&	0					&	0							&	1 - \phi_2
	}.
\end{equation*}

\section{Analytical analysis}
In this section we will investigate the lagging anchor dynamics on a theoretical level. We first briefly discuss the outcome of deterministic learning, and then derive analytical expressions characterizing the properties of stochastic learning in the limit of small noise.

\subsection{Deterministic system}
The deterministic limit of the lagging anchor update is obtained by replacing $X(t)\to x(t)$ and $Y(t)\to y(t)$ in~\eref{fullsystem}, or equivalently by setting $\xi(t)=\chi(t)=0$. The resulting dynamics have been studied previously by Dahl~\cite{Dahl2001,Dahl2002,Dahl2005} and by Butterworth and Shapiro in~\cite{Butterworth2009}.  Dahl considered the case where $\kappa_1 = \kappa_2$, $\mu_1 = \mu_2 = \nu_1 = \nu_2$, and the game is zero-sum, and showed that, provided the learning rate and anchor parameter are small enough and the payoff matrix is invertible, a broad class of mixed equilibrium strategies are asymptotically stable. The authors of~\cite{Butterworth2009} continued this analysis, and determined the region of the parameter space for which the dynamics is stable. We do not repeat the full analysis for the discrete-time system here, as it is straightforward to see that the stability of the Nash point depends on the eigenvalues of the matrix $J$  in~\eref{fullsystemmatrix}. In the interior of the strategy simplexes, the deterministic dynamics is simply given by the linear map $\bzeta(t+1) = J\bzeta(t)$, so that $\bzeta=0$ is a stable fixed point provided all eigenvalues of $J$ have moduli less than one.  In this case, the strategies will converge to the equilibrium point, no matter from what initial condition the dynamics is started. Asymptotically the payoffs to the two players will be those at the equilibrium point, i.e.\ the long-term payoff for each player vanishes in the matching-pennies game.

\begin{figure}
	\centering
	\includegraphics[scale=0.7]{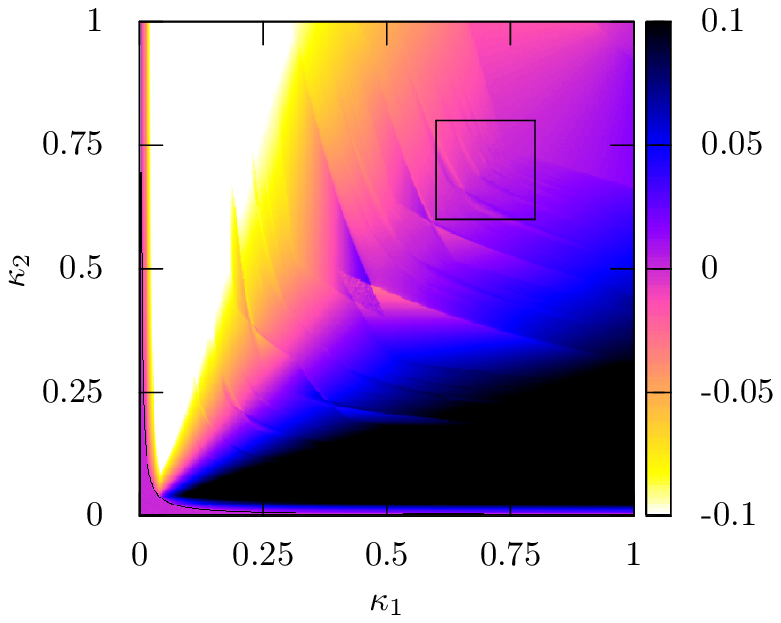}
	\includegraphics[scale=0.7]{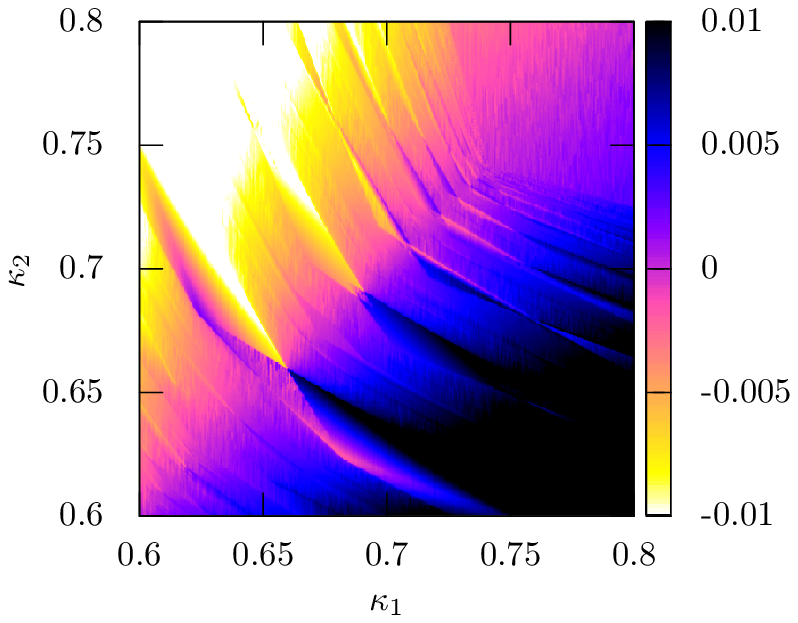}
	\caption{(Colour on-line) Colour maps of the average payoff to player~$1$ for various values of $\kappa_1$ and $\kappa_2$.  The right plot covers the region marked by a square in the left plot, and shows clear fractal behaviour.}
	\label{fractal}
\end{figure}

When the matrix $J$ has one or more eigenvalues outside the unit circle of the complex plane, the strategies spiral outwards until they reach the simplex boundaries.  Then the nonlinearity sets in, effectively clipping the phase space of the system. Numerically iterating the deterministic map we typically observe periodic motion.  Interestingly, we find that if the players are not symmetric (i.e., if they have different learning rates, anchor parameters, or modelling rates), it is possible for one player to consistently achieve a positive payoff. The dependence of this average payoff on the parameters appears to be quite complex, and shows discontinuities in parameter space, as well as fractal-like behaviour, as shown in figure~\ref{fractal}.

\subsection{Stochastic system}
The authors of~\cite{Butterworth2009} determined the form of the covariance of the players' strategies in the continuous limit of the stochastic dynamics, concluding that where the deterministic dynamics is stable, the strategies will typically perform quasicycles about the equilibrium point, with a frequency determined by the eigenvalues of the Jacobian matrix.  We continue this analysis by calculating analytic approximations of the covariances and power spectra of the state variables, and comparing them to results from simulations.

\subsubsection{Assumptions}
The chief barrier to analytically calculating statistics of the lagging anchor system~\eref{fullsystem} is the presence of the truncations that are applied if the strategies leave their allowed intervals.  We assume that the parameters of the algorithm are such that the deterministic dynamics converges, and that the intrinsic noise caused by the limited information is small.  Under these conditions, the strategies spend most of their time near the equilibrium point and so the probability of them reaching the boundaries of the simplexes is negligible, and the nonlinear corrections can be ignored.

\subsubsection{Correlations of the noise terms}
The correlation matrix $D(t) = \avg{\bvarphi(t) \bvarphi(t)^T}$ can be written in block form as
\begin{equation*}
	D(t) = \pmatrix{
		0	&	0												&	0												\cr
		0	&	\phi_1^2 \avg{\xi(t)^2}	&	0												\cr
		0	&	0												&	\phi_2^2 \avg{\chi(t)^2}
	},
\end{equation*}
since the two noise terms $\xi(t)$ and $\chi(t)$ are independent.  The variance of the noise terms $\xi$ and $\chi$ can be calculated from their definitions~\eref{noisedef}, leading to
\begin{equation*}
	\avg{\xi^2} = \frac{1}{4} - x^2, ~~~ \avg{\chi^2} = \frac{1}{4} - y^2.
\end{equation*}

To simplify the calculation of the covariances and power spectra of the strategies, we neglect the quadratic terms in $x$ and $y$ in these expressions.  This assumption can be expected to be accurate close to the Nash point.\footnote{Similar to~\cite{Galla2009,Galla2011,Bladon2011,Gomez2011} this is equivalent to a systematic expansion in the amplitude of the intrinsic noise.}  Under this simplifying assumption, $D(t)$ can be regarded as constant,
\begin{equation*}
	D(t) = D = \frac{1}{4} \pmatrix{
		0	&	0					&	0					\cr
		0	&	\phi_1^2	&	0					\cr
		0	&	0					&	\phi_2^2
	}.
\end{equation*}

\subsubsection{Correlations of the state variables}
We calculate an approximation of the equal-time covariance matrix $C(t)$ of the state vector $\bzeta(t)$,
\begin{equation*}
	C(t) = \avg{\bzeta(t) \bzeta(t)^T}.
\end{equation*}
Using the matrix form~\eref{fullsystemmatrix} of the lagging anchor algorithm and  the fact that $\bvarphi(t)$ and $\bzeta(t)$ are uncorrelated, we have
\begin{equation}\label{covariancerecurrence}
	C(t+1) = J C(t) J^T + D.
\end{equation}

We are interested in the long-term value of $C(t)$, after transients have died away. In this asymptotic regime the components of $C$ are related to the size of the stochastic deviations from the equilibrium point.  If this limiting value exists, it is given by the fixed point $C^*$ of~\eref{covariancerecurrence},
\begin{equation}
	C^* = J C^* J^T + D.
\end{equation}
This matrix equation is easily solved by rearranging it into a linear system~\cite{Kitagawa1977}.  It has a unique positive definite solution whenever the deterministic lagging anchor dynamics has a stable fixed point~\cite{Michel2008}.

The long-term expected payoff $\avg{u}$ received per game by player~$1$ can be calculated given $C^*$, using
\begin{equation*}
\eqalign{
	\avg{u}	 =  4 \avg{x y}  =  4 C^*_{12}.
}
\end{equation*}

\subsubsection{Power spectra}
To check whether the stochastic dynamics display quasicycles, we define a power spectral density matrix for the vector $\bzeta$,
\begin{equation*}
	P(\omega) = \lim_{L \rightarrow \infty} \frac{1}{L} \avg{\hbzeta(\omega, L) \hbzeta(\omega, L)^{\dagger}},
\end{equation*}
where $\dagger$ is the Hermitian conjugate, and the hats denote discrete Fourier transforms,
\begin{equation*}
	\hbzeta(\omega, L) = \sum_{t = 0}^{L-1} \bzeta(t) \exp(-\rmi \omega t).
\end{equation*}
An expression for the power spectra can be obtained by taking the Fourier transform of the lagging anchor recurrence relation~\eref{fullsystemmatrix}, which, for $L\gg 1$, transforms the system into
\begin{equation*}
	(\rme^{\rmi \omega} I - J) \hbzeta(\omega, L) = \hbvarphi(\omega, L),
\end{equation*}
where $I$ is the $6 \times 6$ identity matrix.  Letting $M(\omega) = \rme^{\rmi \omega} I - J$ we obtain
\begin{equation} \label{spectraintermediateeqn}
	P(\omega) = \lim_{L \rightarrow \infty} \frac{1}{L} M(\omega)^{-1} \avg{ \hbvarphi(\omega, L) \hbvarphi(\omega, L)^{\dagger} } M(\omega)^{\dagger^{-1}}.
\end{equation}
If we write the expectation value in this last expression in terms of the noise terms,
\begin{equation*}
	\avg{ \hbvarphi(\omega, L) \hbvarphi(\omega, L)^{\dagger} } = \sum_{t=0}^{L-1} \sum_{s=0}^{L-1} \avg{\bvarphi(t) \bvarphi(s)^T} \rme^{- \rmi \omega (t-s)}.
\end{equation*}
Using the fact
\begin{equation*}
	\avg{\bvarphi(t) \bvarphi(s)^T} = D \delta_{ts},
\end{equation*}
we find
\begin{equation*}
	\avg{ \hbvarphi(\omega, L) \hbvarphi(\omega, L)^{\dagger} } = L D.
\end{equation*}
Substituting this into~\eref{spectraintermediateeqn} gives the result
\begin{equation}
	P(\omega) = M(\omega)^{-1} D M(\omega)^{\dagger^{-1}}.\label{powercalculation}
\end{equation}
This expression is similar to those obtained for population-based models by means of the celebrated van Kampen expansion, see e.g.~\cite{McKane2005, Reichenbach, Bladon}. Studies of intrinsic noise in game learning can also be found in~\cite{Galla2009,Galla2011,Bladon2011,Gomez2011}. Expression~\eref{powercalculation} provides an explicit prediction for the spectra properties of fluctuations about the Nash point (or equivalently a prediction of its correlation function). These predictions can be tested against simulations, using a large finite value of $L$.

\section{Test against simulations}
\subsection{Identical players}
First, we consider the case where the two players have identical update rules (i.e., $\kappa_1 = \kappa_2$, $\mu_1 = \mu_2$, $\nu_1 = \nu_2$, and $\phi_1 = \phi_2$).  We run simulations of the lagging anchor algorithm for the matching pennies game, comparing the results to the analytic calculations discussed above, for various values of the four parameters.

\begin{figure}
	\centering
	\includegraphics[scale=0.75]{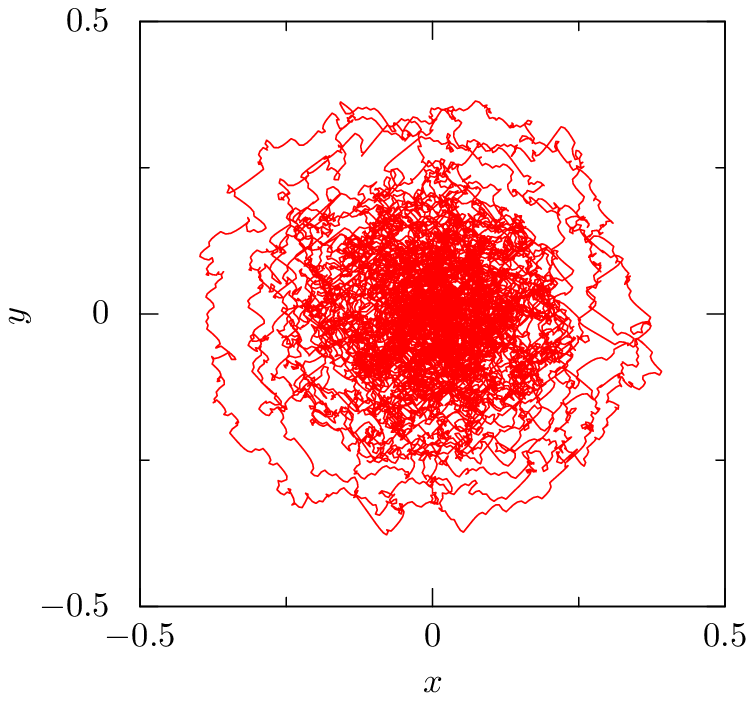}
	\includegraphics[scale=0.75]{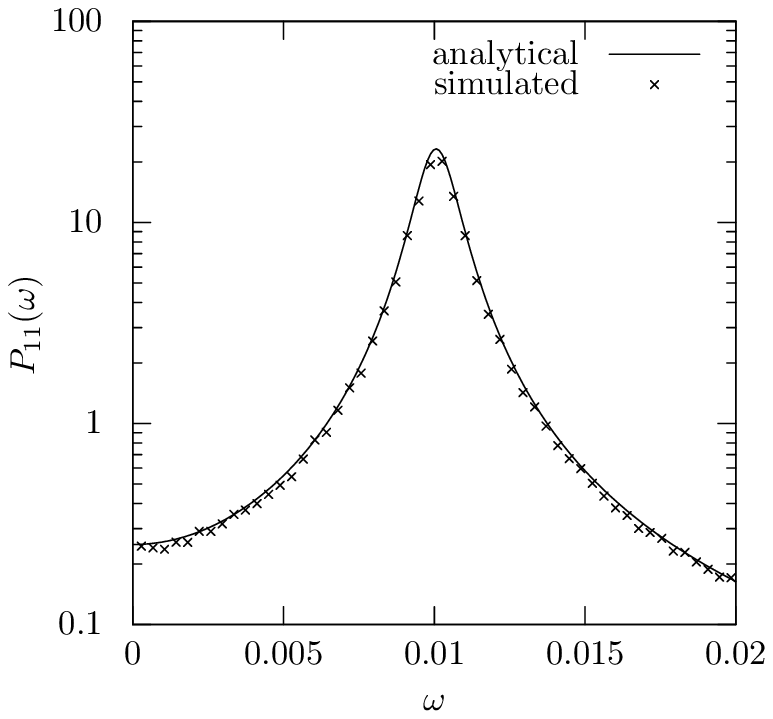}
	\caption{(Colour on-line) {\bf Left:} Phase plot of the strategy coordinates $x$ and $y$ of the two players in a simulation of the lagging anchor algorithm, for the matching pennies game.  The parameters have the values $\kappa_i = 0.005$, $\mu_i = \nu_i = 0.05$, $\phi_i = 0.5$, for which the deterministic dynamics converges to the equilibrium point. {\bf Right:} Corresponding power spectrum of the first player's strategy coordinate $x$.  The solid curve shows the spectrum calculated analytically using~\eref{powercalculation}, while the crosses show a simulated spectrum, averaged over $1000$ realizations of the dynamics ($L=2^{14}$).}
	\label{single_case_stable_phase_plot}
\end{figure}

The left-hand panel of figure~\ref{single_case_stable_phase_plot} shows a phase plot of the players' strategies in one realization, for values of the parameters where the deterministic dynamics is stable.  The right-hand panel shows the associated spectral density of player one's strategy, which shows a single sharp peak, reflecting the existence of quasicycles.  The analytic approximation of the power spectrum is very close to the simulated spectrum, suggesting that the assumptions used to calculate it---i.e., ignoring the restriction of $x$ and $y$ to intervals, and neglecting the quadratic terms in $D(t)$---hold for these values of the parameters.

\begin{figure}
	\centering
	\includegraphics[scale=0.75]{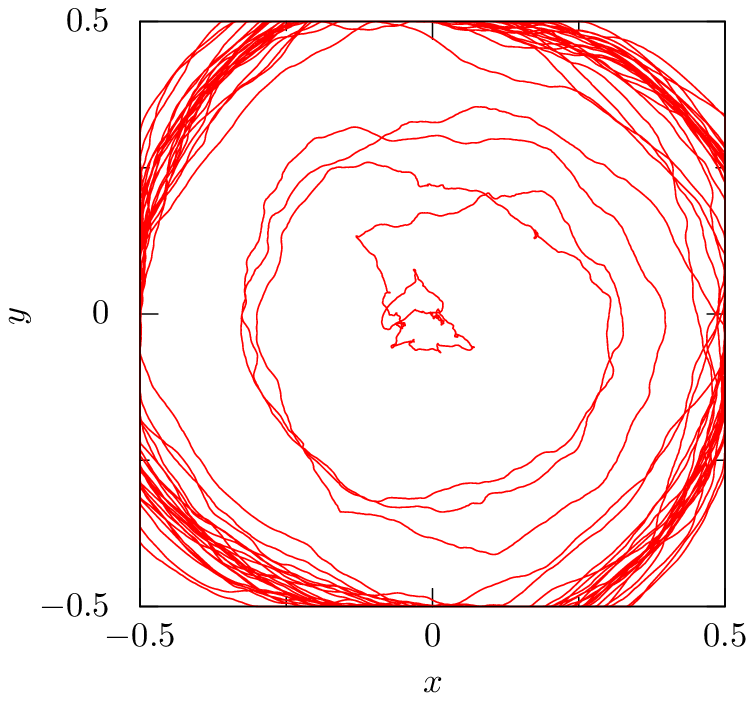}
	\includegraphics[scale=0.75]{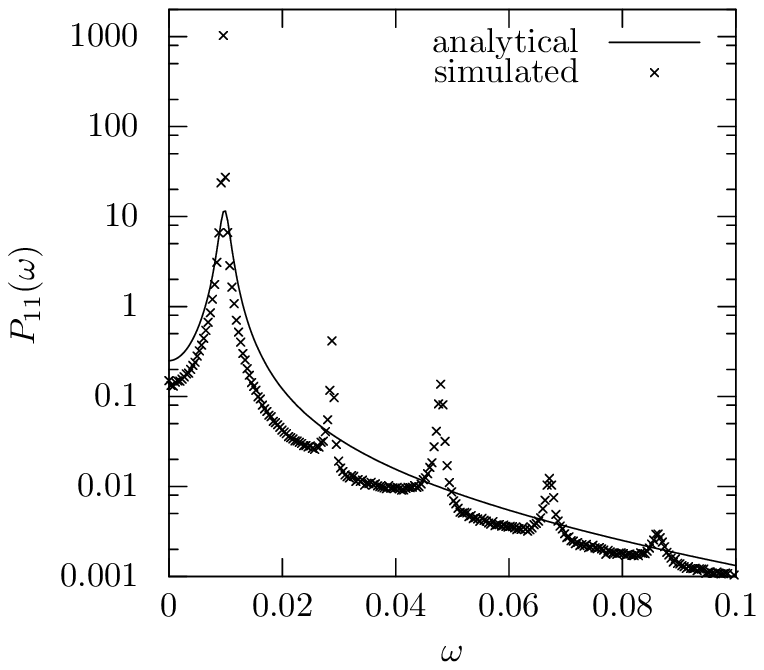}
	\caption{(Colour on-line) {\bf Left:} Phase plot of the strategy coordinates $x$ and $y$ of the two players in a simulation of the lagging anchor algorithm, for the matching pennies game.  The parameters have the values $\kappa_i = 0.005$, $\mu_i = \nu_i = 0.05$, $\phi_i = 0.05$, for which the deterministic dynamics do not converge. {\bf Right:} Corresponding power spectrum of the first player's strategy coordinate $x$.  The solid curve shows the values calculated analytically using~\eref{powercalculation}, while the crosses show a simulated spectrum, averaged over $1000$ realizations of the dynamics.  The sharp peaks in the simulated spectrum are reminiscent of those of a noisy square wave, demonstrating the dominance of the nonlinear truncation of the strategies for these parameters.}
	\label{single_case_unstable_phase_plot}
\end{figure}

In contrast, figure~\ref{single_case_unstable_phase_plot} (left-hand panel) shows a phase plot of simulations for another set of parameters, where the deterministic dynamics is unstable.  The strategies perform a noisy version of the limit cycle found in the deterministic dynamics, generally staying near the boundaries.  The power spectrum of player~$1$'s strategy, shown in the right-hand panel is not in good agreement with the analytic calculation~\eref{powercalculation} as the strategies do not remain close to the equilibrium point.  However, the analytic spectrum correctly predicts the location of the main peak.

Perhaps the most important diagnostic of the effects of changing the parameters is the long-term average variance of the components of the strategies.  This quantifies the size of stochastic oscillations about the equilibrium point, and therefore the success of the algorithm in learning the equilibrium strategies.

\begin{figure}
	\centering
	\includegraphics[scale=0.75]{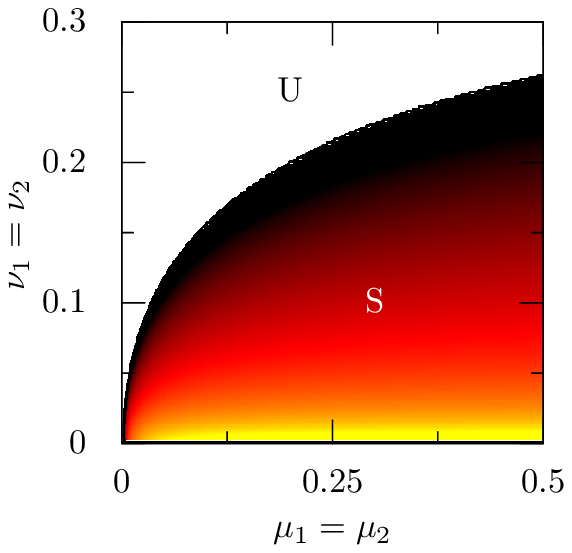}
	\includegraphics[scale=0.75]{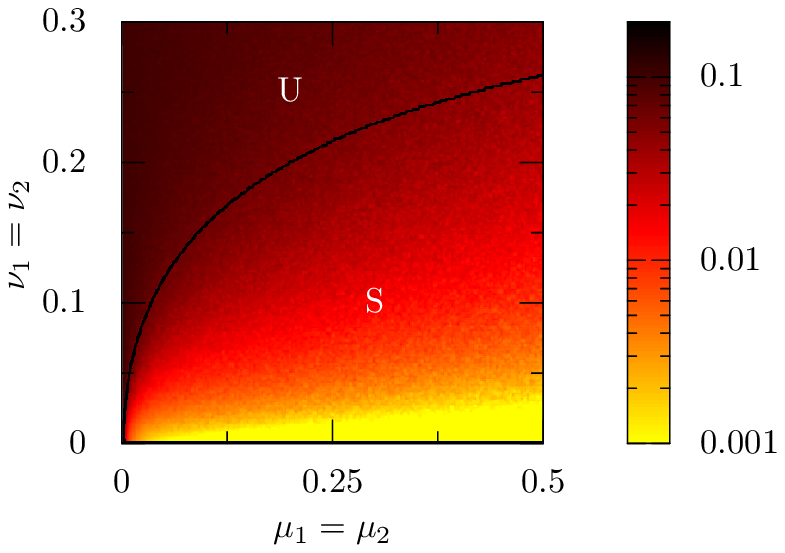}
	\caption{(Colour on-line) Colour maps showing the dependence of the variance of player~$1$'s strategy coordinate $x$ on the anchor parameters $\mu$ and $\nu$, while keeping the other parameters fixed at $\kappa_i = 0.005$, $\phi_i = 0.5$, for the matching pennies game.  The panel on the left shows the analytic approximation, and the right-hand panel the results from simulations (averaged over multiple runs), with the strategies and anchors beginning at the equilibrium point.  In each case, the solid line shows where the deterministic dynamics become unstable.  In the left-hand panel, the variance diverges as the instability region U is approached.}
	\label{vary_mu_nu_variance_maps}
\end{figure}

Figure~\ref{vary_mu_nu_variance_maps} shows the dependence of the size of the oscillations on the anchor parameters $\mu$ and $\nu$.  In general, it appears that in order to have small oscillations, the memory-loss parameter $\nu$ must be small compared to $\mu$.  In other words, the lagging anchors must move slowly and pull the strategies towards them strongly.  The opposite situation, in which the anchors quickly move towards the strategies, but only weakly affect them, is comparable to removing the anchor terms.

Far inside the region where the deterministic dynamics is stable, oscillations are found to be small, and the analytic and simulated variances are very similar.  As the stability line is approached, the analytic approximation of the variance diverges, as the simplex boundaries were ignored in the calculation. The variance measured in simulations approaches a constant value, as the strategies in the simulations are not allowed to leave the bounded intervals on which they are defined.

\begin{figure}
	\centering
	\includegraphics[scale=0.75]{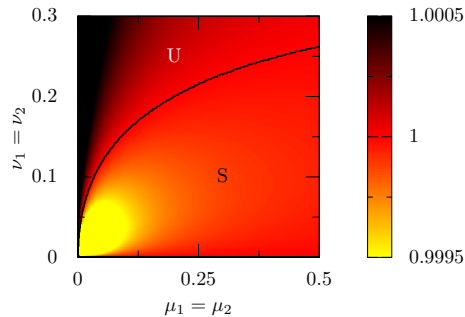}
	\caption{(Colour on-line) Colour map showing the dependence of the largest eigenvalue modulus $\lambda$ of $J$ on the anchor parameters $\mu$ and $\nu$ for the same values of $\kappa$ and $\phi$ as in figure~\ref{vary_mu_nu_variance_maps}.  The solid black curve is the level set for which this modulus is one, and marks the boundary between the stable (S) and unstable (U) behaviours of the deterministic dynamics.}
	\label{vary_mu_nu_eigenvalue_map}
\end{figure}

The long-term variance of the strategies does not completely characterize the behaviour of the algorithm, however.  In the low-$\nu$ region of the plots in figure~\ref{vary_mu_nu_variance_maps}, the anchors will move very slowly, so that if they are not initially close to the equilibrium point, it will take a long time for transients to die away.  The colour map in figure~\ref{vary_mu_nu_eigenvalue_map} shows the largest modulus of the eigenvalues of $J$, which we denote by $\lambda$.  This quantity determines the stability of the deterministic dynamics.  Where $\lambda$ is significantly smaller than one, the deterministic dynamics quickly converge, and so transients in the stochastic dynamics are short-lived.  Where $\lambda$ is close to, but smaller than one, the deterministic dynamics converges slowly, so the transients are long-lasting.  If the lagging anchor algorithm were used in a practical situation, it may be necessary to choose the anchor parameters to strike a balance between having small oscillations and short-lived transients.

\subsection{Two-player learning with non-identical players}
We now turn to situations in which the two players use different values of the parameters $\kappa$, $\mu$, $\nu$, and $\phi$, still using the  matching pennies game.  We keep all but one of the parameters constant, then run simulations for a range of values of the remaining parameter for each player.

\begin{figure}
	\centering
	\includegraphics[scale=0.75]{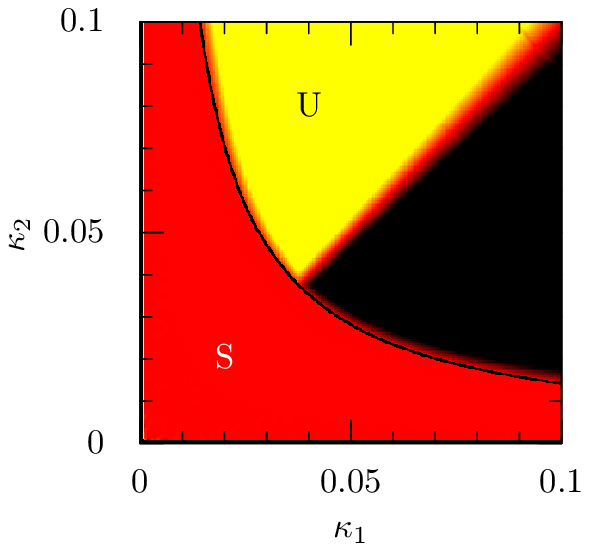}
	\includegraphics[scale=0.75]{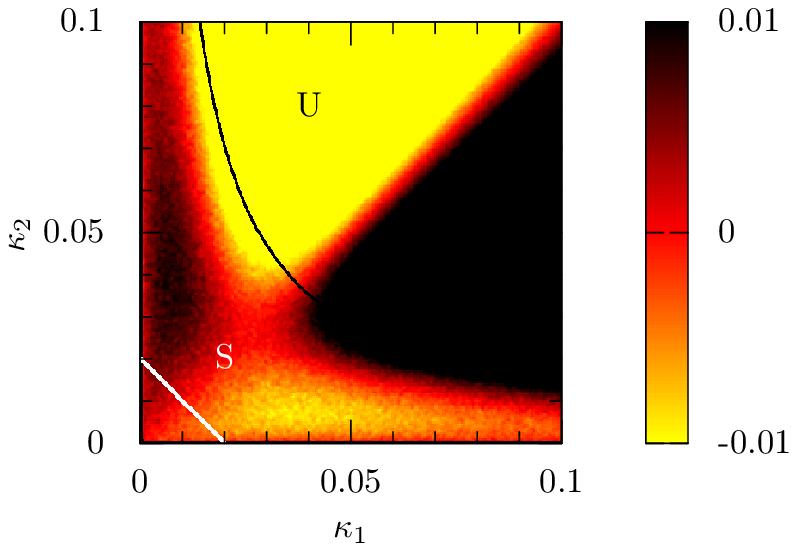}
	\caption{(Colour on-line) Colour maps showing the average payoff to player~$1$ for varying values of $\kappa_1$ and $\kappa_2$, with the other parameters kept fixed at $\mu_i = \nu_i = 0.05$ and $\phi_i = 0.5$.  Left: deterministic dynamics, right: stochastic learning.  The solid black lines mark the boundary between the stable (S) and unstable (U) phases of the deterministic dynamics.  Figure~\ref{cut_kappa1_kappa2} compares the stochastic behaviour along the solid white line $\kappa_1 + \kappa_2 = 0.02$ with analytic predictions.}
	\label{vary_kappa1_kappa2_payoff_maps}
\end{figure}

An interesting quantity to consider is the long-term average payoff to one of the players (recall that we are considering a zero-sum game in which the players receive equal and opposite payoffs).  Figure~\ref{vary_kappa1_kappa2_payoff_maps} shows the dependence of this quantity on the learning rates of the two players, while the other parameters are kept constant, in both the stochastic and deterministic dynamics.  Outside the stability region, the deterministic and stochastic behaviour are similar---the player with the highest learning rate wins.  This is due to the dominant influence of the truncation of the strategies to their intervals.  The structures do not match exactly however---while a player's payoff reduces to zero as the stability region is approached in the deterministic case, it remains nonzero a little way beyond the stability line in the stochastic dynamics---this appears to be because, just inside the stability region, the stochastic quasicycles push the strategies outwards to the boundaries, where they mimic the limit cycles, and payoffs, found in the unstable deterministic case.

Where the deterministic dynamics is stable, it converges to the equilibrium point, where the average payoff is zero, but this behaviour is not seen in the stochastic dynamics, in which a player can still obtain a finite payoff well inside the stability region. Similar effects of increased payoffs to one player driven by intrinsic noise have been observed in learning processes based on noisy replicator dynamics in~\cite{Traulsen2004,Rohl2008}.

\begin{figure}
	\centering
	\includegraphics[scale=0.75]{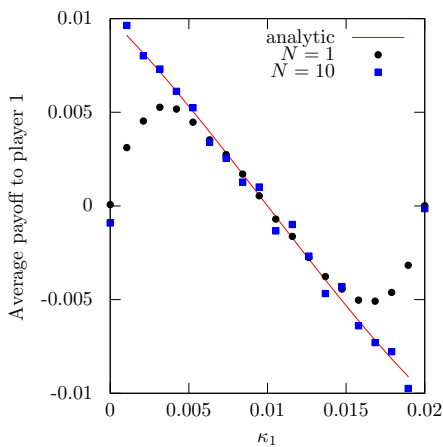}
	\caption{(Colour on-line) Plot showing the payoff to player~$1$ as the learning rates are varied along the line $\kappa_1 + \kappa_2 = 0.02$, with other parameters as in figure~\ref{vary_kappa1_kappa2_payoff_maps}.  The solid line shows the analytic values, squares the simulated values.  Circles show how the simulated values change if there is batching---the players only update their strategies every $N$ games, where in this case $N=10$.  Batching reduces the size of stochastic effects, improving the match with the analytic calculations.}
	\label{cut_kappa1_kappa2}
\end{figure}

Figure~\ref{cut_kappa1_kappa2} compares the analytic values of the average payoff far inside the stability region with numerical simulations. The predictions are accurate near the $\kappa_1 = \kappa_2$ line, but systematic deviations become apparent as $\kappa_1$ or $\kappa_2$ approach zero, as one player's strategies are able to move far away from the equilibrium, so the small-noise approximation becomes inappropriate. In fact adopting a batch learning approach (see~\cite{Galla2009, Galla2011}) increases the range of accuracy of the theory. In batch learning players play `batches' of $N$ observations at each update step, i.e.\ the random variables $X(t)$ and $Y(t)$ in~\eref{fullsystem} are replaced by averages over $N$ independent draws of actions from the players' mixed strategies. This procedure reduces the noise level (which scales as $N^{-1/2}$), and as seen in figure~\ref{cut_kappa1_kappa2} the analytic predictions become more accurate as $N$ is increased.

\subsection{Other games}
The lagging anchor algorithm can easily be applied to other games. Suppose a game has $m$ pure strategies for player~$1$ and $n$ for player two.  Let player~$1$'s strategy be represented by the vector $\bp$, and player~$2$'s by $\bq$, and let $\bp^*$ and $\bq^*$ be the mixed strategies at an equilibrium point.  We introduce strategy coordinates $\bx$ and $\by$ defined by
\begin{equation*}
\eqalign{
	\bp = L\suone \bx + \bp^*, \cr
	\bq = L\sutwo \by + \bq^*,
}
\end{equation*}
where $L\suone$ and $L\sutwo$ are $m \times (m-1)$ and $n \times (n-1)$ matrices, respectively, of the form
\begin{equation*}
	L^{(1)} = \left(\begin{array}{c} I_{m-1} \\ -\bone_{m-1}^T\end{array}\right), ~~ L^{(2)} = \left(\begin{array}{c} I_{n-1} \\ -\bone_{n-1}^T\end{array}\right),
\end{equation*}
where $I_k$ denotes an identity matrix of size $k \times k$, and $\bone_k$ a column vector of ones of length $k$.
\begin{figure}
	\centering
	\includegraphics[scale=0.6]{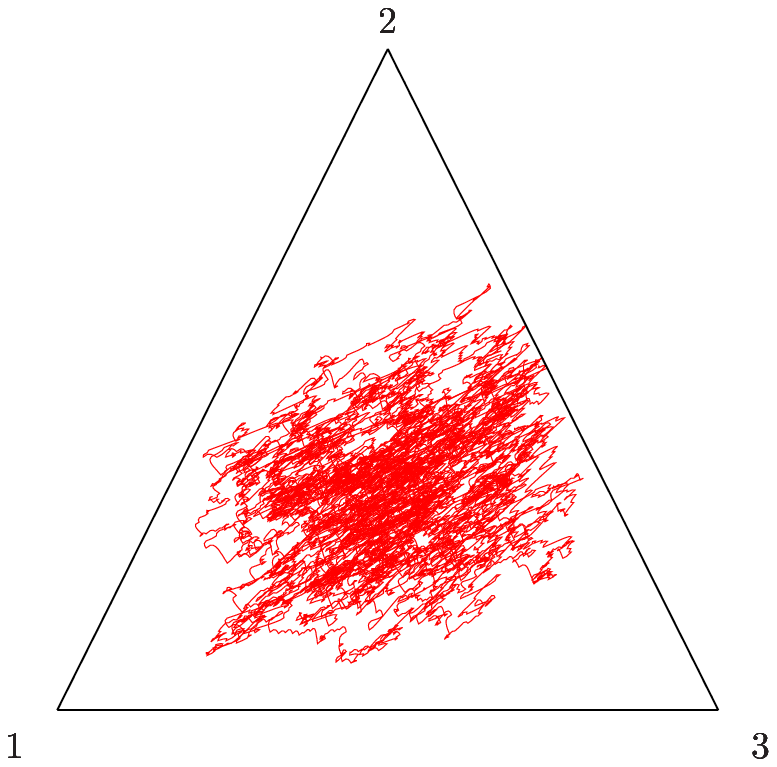}
	\includegraphics[scale=0.6]{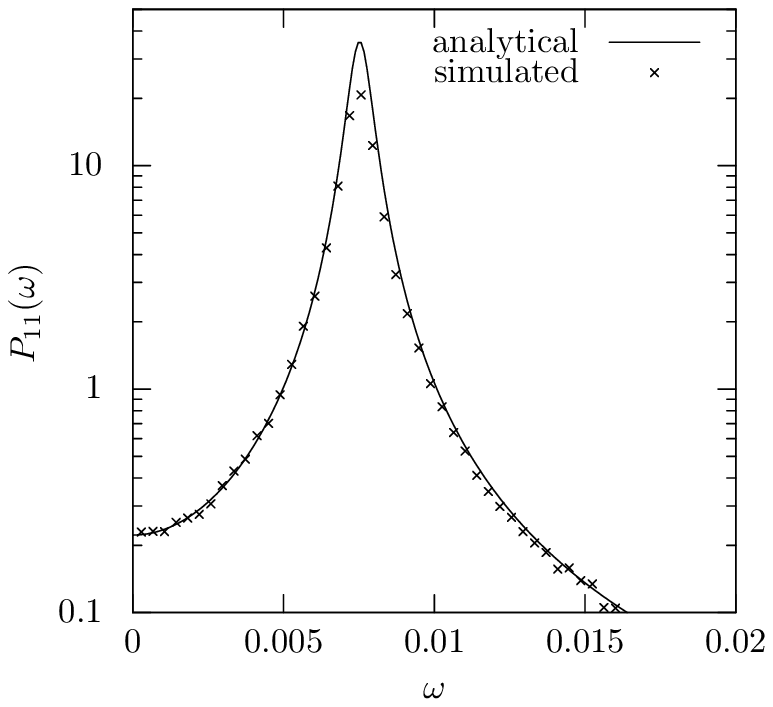}
	\caption{(Colour on-line) Left: Ternary (or barycentric) plot of the components of player~$1$'s strategy in a realization of the lagging anchor algorithm, for the rock-paper-scissors game.  The parameters have the values $\kappa_i = 0.005$, $\mu_i = \nu_i = 0.05$, $\phi_i = 0.5$, for which the deterministic dynamics converges.  Right: Power spectrum of a component of player~$1$'s strategy. Solid line: theory, markers: simulation (averaged over $1000$ simulations).}
	\label{rps_single_case_stable_phase_plot}
\end{figure}
Let the players' payoff matrices be $E\suone$ and $E\sutwo$, respectively.  Then the expected payoffs $u_1$ and $u_2$ can be written
\begin{equation}\label{eqgrad}
\eqalign{
	u_1 =\bx^T A\suone \by + \bp^{*^T} E\suone L\sutwo \by + \bp^{*^T} E^1 \bq^*, \cr
	u_2 = \bx^T A\sutwo \by + \bq^{*^T} E\sutwotrans L\suone \bx + \bp^{*^T} E^2 \bq^*,
}
\end{equation}
where $A\suone = L\suonetrans E\suone L\sutwo$ and $A\sutwo = L\suonetrans E\suone L\sutwo$.  Terms of the form $\bx^T{L^{(1)}}^TE^{(1)}\bq^*$ and $\by^T{L^{(2)}}^TE^{(2)}\bp^*$ vanish for Nash equilibria in the interior of strategy space, and we will only consider such cases here. The expressions in~\eref{eqgrad} allow the payoff gradients required for the gradient ascent algorithm to be written in a simple form,
\begin{equation*}
\eqalign{
	\nabla_{\bx} u_1 = A\suone \by, \cr
	\nabla_{\by} u_2 = A\sutwo \bx,
}
\end{equation*}
so that the full lagging anchor system~\eref{fullsystem} becomes
\begin{equation*}
\eqalign{
	\bx(t+1) = \bx(t) + A\suone \tby(t) + \mu_1 (\obx(t) - \bx(t))  \cr
	\by(t+1) = \by(t) + A\sutwo \tbx(t) + \mu_2 (\oby(t) - \by(t)) \cr
	\obx(t+1) = \obx(t) + \nu_1 (\bx(t) - \obx(t)) \cr
	\oby(t+1) = \oby(t) + \nu_2 (\by(t) - \oby(t)) \cr
	\tbx(t+1) = \tbx(t) + \phi_1 (\bX(t) - \tbx(t)) \cr
	\tby(t+1) = \tby(t) + \phi_2 (\bY(t) - \tby(t)).
}
\end{equation*}

Now, instead of intervals, the players' strategies are confined to standard simplexes---that is, sets of vectors whose elements are nonzero and sum to one.  If the update rules take the strategies outside their simplexes, they must be mapped back to the nearest point in the simplex---this is a `convex projection', described in detail by Michelot~\cite{Michelot1986}.

The calculation of the approximations of the covariances and power spectra follow in a similar manner to those for a two-action game.  As an example, a phase plot of the lagging anchor dynamics for the `rock-paper-scissors' game with the same values of the constant parameters as that used for the matching pennies game in figure~\ref{single_case_stable_phase_plot} is shown in figure~\ref{rps_single_case_stable_phase_plot}, with a corresponding power spectrum.  The dynamics is closely comparable to those seen for matching pennies, and the approximation of the power spectrum is similarly accurate.

\section{Conclusions}
Intrinsic noise appears in the lagging anchor dynamics when the players do not know their opponents' strategies.  The properties of these fluctuations can be calculated analytically to good approximation, when the deterministic dynamics is stable.  In this case, the noise can have a significant effect on the dynamics, inducing quasicycles, similar to those driven by demographic noise in population-based models~\cite{McKane2005, Reichenbach, Bladon}. See also~\cite{Galla2009,Galla2011,Bladon2011,Gomez2011} for studies of intrinsic noise in game learning. In our work we have considered two-player learning in the context of the matching pennies game, and the well-known rock-paper-scissors game. We are able to predict the magnitude of stochastic fluctuations, and to determine the spectral properties of quasicycles analytically, in good agreement with simulations. When the players are asymmetric, i.e.\ when they use different parameters for their respective lagging anchor adaptation, one of them can take advantage of the resulting quasicycles, achieving a positive average payoff.  While we limit the present work to an analysis of the above two-player games, the analytic framework developed here is very general, and can be applied to any game learning system in which the strategies tend to stay close to an equilibrium point. Convergence to Nash points is a key objective with which many adaptation mechanisms in machine learning have been designed, and as we show in our work intrinsic noise due to imperfect sampling can seriously affect whether or not this objective is achieved. The formalism presented in this paper provides a systematic approach with which to estimate deviations from convergence, and we expect that it can be applied to a large class of machine-learning algorithms, beyond the lagging anchor scheme we have discussed here.

\ack{JBTS is supported by an EPSRC studentship. TG is grateful for funding by the Research Councils UK (RCUK reference EP/E500048/1),  JLS acknowledges funding by EPSRC (grant EP/E050441/1).}

\section*{References}

\end{document}